\documentclass[10pt,conference,compsocconf,letterpaper]{IEEEtran}
\pdfoutput=1
\usepackage{graphicx}
\usepackage{subfigure}
\usepackage[linesnumbered,ruled,vlined]{algorithm2e}
\usepackage{amssymb}
\usepackage{amsmath}
\usepackage{makecell}
\usepackage{array}
\usepackage{url}
\usepackage{flushend}

\begin{document}

\title{A Privacy-Preserving QoS Prediction Framework \\for Web Service Recommendation\vspace{-2ex}}

\author{\IEEEauthorblockN{Jieming Zhu,~Pinjia He,~Zibin Zheng,~Michael R. Lyu\vspace{0.5ex}}
\IEEEauthorblockA{Shenzhen Research Institute, The Chinese University of Hong Kong, Shenzhen, China\\
Department of Computer Science and Engineering, The Chinese University of Hong Kong, Hong Kong\\
\textit{\{jmzhu, pjhe, zbzheng, lyu\}@cse.cuhk.edu.hk}\vspace{1ex}}
}

\maketitle

\begin{abstract}
QoS-based Web service recommendation has recently gained much attention for providing a promising way to help users find high-quality services. To facilitate such recommendations, existing studies suggest the use of collaborative filtering techniques for personalized QoS prediction. These approaches, by leveraging partially observed QoS values from users, can achieve high accuracy of QoS predictions on the unobserved ones. However, the requirement to collect users' QoS data likely puts user privacy at risk, thus making them unwilling to contribute their usage data to a Web service recommender system. As a result, privacy becomes a critical challenge in developing practical Web service recommender systems. In this paper, we make the first attempt to cope with the privacy concerns for Web service recommendation. Specifically, we propose a simple yet effective privacy-preserving framework by applying data obfuscation techniques, and further develop two representative privacy-preserving QoS prediction approaches under this framework. Evaluation results from a publicly-available QoS dataset of real-world Web services demonstrate the feasibility and effectiveness of our privacy-preserving QoS prediction approaches. We believe our work can serve as a good starting point to inspire more research efforts on privacy-preserving Web service recommendation.
\end{abstract}


\section{Introduction}\label{sec:intro}
Web services are self-contained units of software functionalities (e.g., retrieving currency exchange rates) delivered over the Internet for users to build composite Web applications. Recent advances in cloud computing enable on-demand service delivery and promote the rapid growth of service markets, where more and more Web services are expected to become available. Whereas the abundance of Web services meets the various needs of different users (i.e., Web application providers), it also poses a significant challenge in selecting among a large number of similar services~\cite{KleinIH12}. In this context, Web service recommendation~\cite{ZhengMLK09, ZhangSLG14, ChenZYL14} that aims to help users quickly find desirable services has become a hot research issue in the area of service computing in recent years. 

Effective service recommendation needs to fulfil both functional and non-functional requirements of users. While functional requirements focus on what a service does, non-functional requirements are concerned with the quality of service (QoS), such as response time, throughput, and failure probability, etc. QoS plays an important role in Web service recommendation, according to which similar services can be ranked and selected for users. Service invocations usually rely on the Internet for connectivity and are heavily influenced by the dynamic network conditions. Therefore, users at different locations typically observe different QoS values even on the same Web service. To enable personalized Web service recommendation, QoS evaluation from user side is desired. However, it is a challenge to acquire user-perceived QoS values of all the services because each user only has observed QoS values on a few used services. It is also impractical for each user to actively measure these QoS values due to the expensive overhead of invoking a large number of services.

To address this issue, collaborative QoS prediction has recently been proposed, and becomes a key step to QoS-based Web service recommendation. By applying collaborative filtering (CF) techniques~\cite{SuK09} that are widely used in commercial recommender systems, unknown QoS can be predicted based on historical usage data collected from users, eliminating the need of additional service invocations. In other words, users can contribute their historical QoS data on the services they have used and receive prediction results on the QoS values of the services that they have never used before. In recent literature, a number of collaborative filtering approaches have been proposed for QoS prediction. Among them, neighbourhood-based CF approaches (e.g., UIPCC~\cite{ZhengMLK09}) leverage the similarity between users and/or the similarity between services calculated on the observed QoS data for unknown QoS prediction. Model-based approaches (e.g., PMF~\cite{ZhengMLK13}, EMF~\cite{LoYDLW12}) fit the observed QoS data with a pre-defined model (e.g., low-rank matrix factorization), and then utilize the trained model for QoS prediction. Recent studies have shown that these approaches achieve high accuracy of QoS predictions and yield encouraging results on Web service recommendation.

Despite the potential benefits provided by Web service recommender systems, a major impediment to the practical deployment of such systems lies in their threats to user privacy. To receive effective recommendations, users are required to supply their observed QoS values. However, there is currently no policy to protect users from privacy issues. Malicious recommender systems, for example, may abuse the data, infer private information from the data, or even resell the data to a competing user for profits~\cite{NikolaenkoIWJTB13}. Even if the recommender system is not malicious, an unintentional leakage of such data can expose users to a broad set of privacy issues (e.g., QoS data may reveal the underlying application configurations). This is why application providers are not willing to disclose their private usage data to the public or a third party. Such privacy threats limit the QoS data collection from users and hence degrade the accuracy of Web service recommendation. To encourage broader user participation, it is desired to consider privacy-preserving approaches for Web service recommendation that can be made without revealing private user data. Encryption is a straightforward way to achieve privacy. However, encryption techniques usually involve large computational overhead and typically work for distributed collaborative filtering problems (e.g., homomorphic encryption used in~\cite{Canny02}) where multi-party communication is necessary. This is inapplicable to our problem because user-user communication is infeasible.

In this paper, we propose a simple yet effective privacy-preserving framework for QoS-based Web service recommendation. Specifically, users are enabled to obfuscate their private data by data randomization techniques~\cite{PolatD03} before they expose the data to a recommender system. In this way, the recommender system can only collect obfuscated QoS data from users, and thus reduce the risk to expose user privacy. Our privacy-preserving framework is generic and can be applied to both the neighbourhood-based collaborative filtering approach, i.e., UIPCC~\cite{ZhengMLK11}, and the model-based collaborative filtering approach, i.e., PMF~\cite{ZhengMLK13}, which are two most common QoS prediction approaches in recent literature. We further revamp these two existing QoS prediction approaches based on our framework, and develop their corresponding privacy-preserving variants: {P-UIPCC} and {P-PMF}. We evaluate these approaches on WS-DREAM dataset~\cite{ZhengZL14}, a publicly-available QoS dataset that has been widely employed for QoS prediction evaluation in the literature. The experimental results show that while preserving user privacy, our proposed approaches (P-UIPCC and P-PMF) can still attain decent prediction accuracy with comparision to the baseline approaches (UEAN and IMEAN) and the counterpart approaches (UIPCC and PMF). We also show the tradeoff between the achieved prediction accuracy and the preserved user privacy. For reproducibility, we release the source code and detailed evaluation results on our project page\footnote{\url{http://wsdream.github.io/PPCF}}. 

In summary, our paper makes the following contributions:
\begin{itemize}
 \setlength{\itemsep}{0.5ex}
  \item This is the first work to cope with the privacy concerns for QoS-based Web service recommendation.
  \item We propose a simple yet effective privacy-preserving framework, and further develop two representative privacy-preserving QoS prediction approaches, P-UIPCC and P-PMF, under this framework.
  \item We conduct experiments on a real-world large-scale QoS dataset of Web services to evaluate the effectiveness of privacy-preserving QoS prediction approaches.
\end{itemize}

The remainder of this paper is organized as follows. Section~\ref{sec:background} introduces the background and related work. Section~\ref{sec:framework} presents the framework of privacy-preserving Web service recommendation. Then we describe the detailed QoS prediction approaches in Section~\ref{sec:approach}, and report the evaluation results in Section~\ref{sec:experiment}. Finally, we conclude this paper in Section~\ref{sec:conclusion}.

\section{Background and Related Work}\label{sec:background}
In this section, we introduce the background of QoS-based Web service recommendation and review two representative collaborative filtering approaches used for QoS prediction. We then discuss the privacy issues and the key techniques for privacy preservation in related work.

\subsection{QoS-based Web Service Recommendation}
\begin{figure}[t]
  \centering
  \includegraphics[width=0.48 \textwidth]{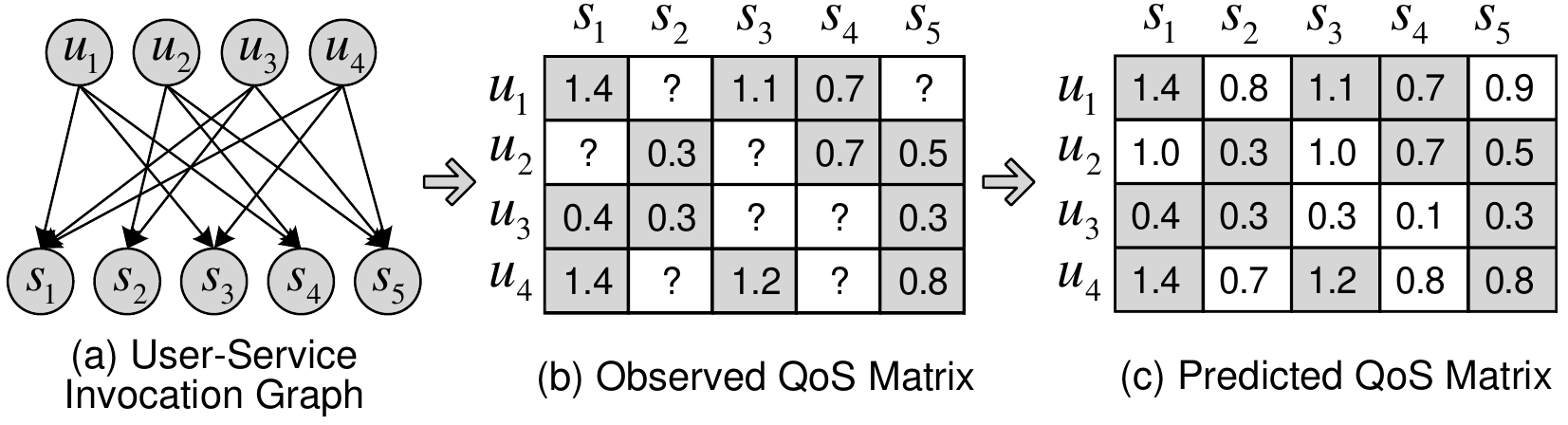}\\
  \vspace{-1ex}
  \caption{An Illustrative Example of QoS Prediction}\label{fig:cfexample}
  \vspace{-2ex}
\end{figure}

QoS-based Web service recommendation has recently attracted much attention from the service community, for providing a promising way to help users select high-quality services out of all the candidate services according to the user-perceived QoS values. Because it is prohibitively expensive or even infeasible for a user to acquire all the QoS values of the candidate services, the key of QoS-based Web service recommendation is to enable accurate QoS predictions.

Collaborative filtering (CF)~\cite{SuK09} has been widely used in commercial recommender systems (e.g., movie recommendation in Netflix, item recommendation in Amazon) for rating prediction, where the observed user ratings are leveraged to learn user preferences on the unrated movies or items and further make predictions on the unknown ratings. In recent literature, CF has been suggested as a promising approach to QoS prediction (e.g.,~\cite{ZhengMLK09, ZhangSLG14, ChenZYL14}). As with the user-movie rating matrix collected in a movie recommender system, users invoking services can produce a user-service QoS matrix with respect to each QoS attribute. We denote a QoS matrix by $R$, whose entry $R_{ij}$ represents the observed QoS value (e.g., response time) of user $u_i$ invoking service $s_j$. Fig.~\ref{fig:cfexample}(b) illustrates a QoS matrix with four users ($u_1$, ..., $u_4$) and five services ($s_1$, ..., $s_5$), produced by the user-service invocation graph in Fig.~\ref{fig:cfexample}(a). In practice, the QoS matrix is very sparse (i.e., most of the entries are unknown), since each user usually invokes only a few services. As shown in Fig.~\ref{fig:cfexample}(b), the grey entries are observed QoS values (e.g., $R_{11} = 1.4$) and the blank entries are unknown QoS values (e.g., $R_{12} =$ ?). As a result, the QoS prediction problem can be modelled as a collaborative filtering problem. Fig.~\ref{fig:cfexample}(c) shows the predicted QoS matrix from the observed QoS matrix in Fig.~\ref{fig:cfexample}(b), where the unknown values are approximately reconstructed. 

Specifically, two types of CF approaches have been studied for QoS prediction of Web services in recent literature:  


\subsubsection{Neighbourhood-based Collaborative Filtering}
This type of CF approaches use the observed QoS data to compute the similarity values between users or services, and further leverage them for QoS prediction. Typical examples include user-based approaches (e.g., UPCC~\cite{ShaoZWZXM07}) that leverage the QoS information of similar users for prediction, item-based approaches (e.g., IPCC~\cite{SarwarKKR01}) that employ the QoS information of similar items (i.e., services) for prediction, and their hybrids (e.g., UIPCC~\cite{ZhengMLK09, ZhengMLK11}) that combine user-based and item-based approaches together for accuracy improvement. These approaches are easy to implement, but they fail to deal with the data sparsity problem, which limits their performance in practice. 

\subsubsection{Model-based Collaborative Filtering}
Model-based CF approaches provide a predefined model to fit the observed QoS data, and then the trained model can be  used to predict the unknown QoS values. Matrix factorization (e.g., PMF~\cite{SalakhutdinovM07}) is one of the most popular model-based CF approaches, which was first introduced to address the QoS prediction problem in~\cite{LoYDLW12}. Matrix factorization model handles the sparsity problem well and usually achieves better performance than neighbourhood-based approaches.

In this paper, we mainly look into two QoS prediction approaches, UIPCC~\cite{ZhengMLK09} and PMF~\cite{SalakhutdinovM07}. They are representatives of the two types of CF approaches respectively and serve as a basis to develop many more sophisticated approaches. For example, some studies such as CloudPred~\cite{ZhangZL11}, NIMF~\cite{ZhengMLK13}, and LN-LFM~\cite{YuLXY14} integrate neighbourhood-based and model-based CF approaches, while some others suggest to leverage additional context information such as location information~\cite{ChenZYL14} and time information~\cite{YuH14, ZhuHZL14} for improving prediction accuracy. Our work focuses on providing a privacy-preserving QoS prediction framework. Therefore, the studies on how to build more sophisticated models for accuracy improvement are orthogonal to our work and fall outside the scope of this paper.

\subsection{Privacy Issues}
Privacy is an important issue that has raised particular concerns among many research areas. In the following, we review the privacy studies related to our work. 

\subsubsection{Privacy in Service Computing}
In service computing, applications are typically built by composing Web services offered by different service providers. User information often needs to be shared across the providers to fulfil an overall application task. This can raise privacy issues between users and service providers when the selected Web services for composition have privacy policies that are not compliant with users' privacy requirements. In this regard, privacy-aware Web service selection and composition (e.g.,~\cite{XuVSR06, SquicciariniCK11, CostantePZ13, TbahritiGMM14}) have been studied. For example, Costante et al.~\cite{CostantePZ13} propose an approach to rank the candidate Web services with respect to the privacy level they offer. Tbahriti et al.~\cite{TbahritiGMM14} further provide a mechanism to verify and negotiate privacy constraints between users and service providers to enable privacy-compatible service composition. Different from these studies, our work aims to address privacy issues for Web service recommendation. 

\subsubsection{Privacy in Recommender Systems}
In recommender systems~\cite{RamakrishnanKMGK01}, users want to gain useful recommendations without compromising their privacy. To achieve so, a variety of privacy-preserving collaborative filtering approaches~\cite{BilgeKYGP13} have been proposed by using techniques such as randomization~\cite{PolatD03}, cryptography~\cite{NikolaenkoIWJTB13}, anonymization~\cite{HoensBSC13}, and so on. Privacy is also of vital importance to the realization of QoS-based Web service recommendation, where users might not be willing to disclose their private usage data. However, there is currently a lack of studies on how to cope with the privacy issues for QoS-based Web service recommendation. Existing privacy-preserving CF approaches are not directly applicable because of the unique challenges posed by Web service recommendation. For example, most of these approaches (e.g.,~\cite{Canny02a, Canny02, ZhanHWHLW10}) require multi-party or peer-to-peer collaboration between users, which is inapplicable to service users. To bridge this gap, our paper makes the first attempt to build a privacy-preserving QoS prediction framework for Web service recommendation. 


\section{Framework of Privacy-Preserving Web Service Recommendation}\label{sec:framework}
\begin{figure}[t]
  \centering
  \includegraphics[width=0.45 \textwidth]{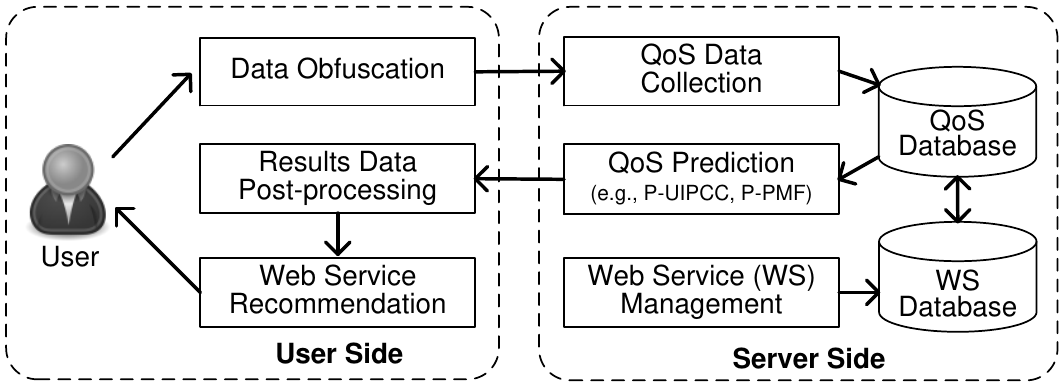}\\
  \vspace{-1ex}
  \caption{Framework of Privacy-Preserving Web Service Recommendation}\label{fig:framework}
  \vspace{-2ex}
\end{figure}

Fig.~\ref{fig:framework} presents our privacy-preserving Web service recommendation framework. The workflow of this framework can be separated into two parts executed at user side and server side respectively. At user side, the observed QoS data of each user undergo a data obfuscation process in order to protect user privacy as well as preserve the information required for performing collaborative QoS prediction. The obfuscated user data are then submitted to the server for QoS prediction. After receiving the prediction results from the server, a post-processing step is performed to recover the obfuscated results to the true QoS prediction values. At last, according to the recovered QoS values, candidate Web services can be ranked and recommended for the user. On the other hand, at server side, obfuscated QoS data are collected from different users in a collaborative way, through which a obfuscated QoS matrix can be acquired and stored in a QoS database. QoS prediction is then performed on the obfuscated QoS matrix by using our proposed privacy-preserving techniques such as P-UIPCC and P-PMF. At the same time, a list of the available Web services is maintained at a Web service database, which allows for service ranking and recommendation for the users. 


User privacy is preserved by our framework because: 1) For each user, user data are obfuscated before being submitted to the server, and the obfuscation settings are only known to the user itself; 2) For the server, collaborative QoS prediction is performed based solely on the obfuscated user data, whereby user-observed real QoS values cannot be inferred. In this way, our framework enables users with greater control on their private usage data and less dependence on the server for privacy preservation. The privacy-preserving framework is generic such that both of the representative QoS prediction approaches (i.e., UIPCC and PMF) can work well without the need of significant modifications.

\section{QoS Prediction Approach}\label{sec:approach}
The above framework enables data obfuscation for preserving privacy, but also poses a challenge in accurate QoS prediction. In this section, we describe the data obfuscation process in detail, and then extend two representative QoS prediction approaches (UIPCC and PMF) into their privacy-preserving variants (P-UIPCC and P-PMF) accordingly. 

\subsection{Data Obfuscation}
The need for privacy preservation has led to the development of a number of data obfuscation techniques, such as data randomization~\cite{PolatD03}, data encryption~\cite{NikolaenkoIWJTB13}, data anonymization~\cite{Klosgen95}. Due to the sparse nature of our data, in this paper, we make use of data randomization~\cite{PolatD03}, a simple yet effective way to obfuscate the data. 

The basic idea of data randomization is to add a random value (i.e., noise) to the true value so that the resulting value becomes disguised. In this way, when the obfuscated QoS data undergo further processing, user information regarding real QoS values can be preserved. Fortunately, although each individual QoS value becomes disguised, we find that some approximate computations (e.g., scalar product) on the aggregated data of users can still be done with decent accuracy. 

To make it clear, we now describe the scalar product property of data randomization~\cite{PolatD03} in detail. Let $a = (a_1, ..., a_n)$ and $b = (b_1, ..., b_n)$ be true vectors with a mean of zero. We obfuscate these vectors as $a' = a + \epsilon$ and $b' = b + \delta$, where $\epsilon = (\epsilon_n, ..., \epsilon_n)$ and $\delta = (\delta_n, ..., \delta_n)$ are random noises generated from a uniform distribution in $[-\alpha, \alpha]$. Next, we show that the scalar product between $a$ and $b$ can be approximated by using the obfuscated vectors $a'$ and $b'$: i.e., $\mathbf{a'b' \approx ab}$. To this end, we have 
\begin{equation}
\small
a'b' =  \sum_{i=1}^n(a_i + \epsilon_i)(b_i + \delta_i) = \sum_{i=1}^n(a_ib_i + a_i\delta_i + b_i\epsilon_i + \epsilon_i\delta_i). \nonumber
\vspace{-0.5ex}
\end{equation}
Because $a$ and $\delta$ are independent vectors and each has a zero mean, we have $\sum_{i=1}^na_i\delta_i \approx 0$. Likewise, we have $\sum_{i=1}^nb_i\epsilon_i \approx 0$, and $\sum_{i=1}^n\epsilon_i\delta_i \approx 0$. Hence, we derive the following approximation: 
\vspace{-1ex}
\begin{equation} \label{equ:scalarproduct}
\small
a'b' \approx \sum_{i=1}^na_ib_i = ab.
\vspace{-2ex}
\end{equation}

With this observation, we find that data randomization can potentially preserve user privacy as well as the usability of the data for collaborative analysis. Therefore, it is appealing to study how to apply this data obfuscating technique to performing collaborative QoS prediction in a privacy-preserving way. To achieve this goal, we propose a two-step data obfuscation procedure for QoS data processing. We emphasize that, as shown in our framework in Fig.~\ref{fig:framework}, each user performs data obfuscation individually at user side before contributing the QoS data to the server.

\subsubsection{\textbf{Z-score normalization}}
To facilitate better randomization of the data, we perform z-score normalization on the observed QoS data as the first step. Z-score normalization is a standard normalization method to adjust the data average and data variance. The normalized data have a zero mean and unit variance. More specifically, for user $u$, we denote $R_{u} = (R_{u1}, ..., R_{um})$ as a vector of observed QoS values on $m$ Web services. $R_{us} > 0$ indicates that user $u$ has invoked service $s$; otherwise, $R_{us}=0$. We compute the mean (${\bar R_u}$) and standard deviation ($\sigma_u$) of this QoS vector $R_{u}$:
\begin{equation}\label{equ:sigma}
\small
\bar R_u = \sum\nolimits_{s \in I_u}{R_{us}}/|I_u|,~~\sigma_u = \sqrt{\sum\nolimits_{s \in I_u}(R_{us} - \bar {R}_u)^2 / |I_u|},
\vspace{-0.5ex}
\end{equation}
where $I_u = \{s~|~R_{us} > 0\}$ denotes the set of Web services that has been invoked by user $u$. Then z-score normalization is performed on the QoS values with the following equation:
\begin{equation}\label{equ:z-score}
\small
r_{us} = (R_{us} - {\bar R}_u) / \sigma_u.
\vspace{-0.5ex}
\end{equation}
The normalization step results in a zero-mean data vector that is well suited for the following data randomization process.

\subsubsection{\textbf{Data Randomization}}
As the second step, we perform randomized perturbation on the normalized QoS vector by:
\begin{equation} \label{equ:data-random}
\small
r'_{us} = r_{us} + \epsilon_{us},
\vspace{-0.5ex}
\end{equation}
where $\epsilon_{us}$ is a random value generated from a specified distribution, for example, uniform distribution in [$-\alpha$, $\alpha$]. Especially when $\alpha=0$, the overall data obfuscation process reduces to a z-score normalization. We further study the effect of different distributions (e.g., uniform distribution, Gaussian distribution) of random noises on QoS prediction accuracy in Section~\ref{sec:impact_random}.

After data obfuscation, users can submit their obfuscated QoS data to the server. Given $n$ users and $m$ services, the server can collect a QoS matrix denoted as $r' \in \mathbb{R}^{n \times m}$ with each entry ($r'_{us}$) being obtained via Equ. (\ref{equ:data-random}). Since such data obfuscation process is performed at user side, the private information such as $\bar{R}_u$ and $\sigma_u$ are kept at user side. As a result, the server cannot infer the true QoS values of the users, and user privacy is preserved.

Next, we will show how we extend the two representative approaches (UIPCC and PMF) to perform privacy-preserving QoS prediction based on the obfuscated QoS matrix $r'$. Note that UIPCC and PMF have been carefully reported in the related work~\cite{ZhengMLK11, ZhengMLK13}, so we do not intend to provide the original descriptions but the necessary extensions from them.

\subsection{Privacy-Preserving UIPCC (P-UIPCC)}
UIPCC (a.k.a. WSRec), first proposed in~\cite{ZhengMLK09}, has been a widely-studied QoS prediction approach. The key of UIPCC is to compute the similarity between users and the similarity between services, after which QoS values contributed by similar users and similar services can be leveraged to compute the prediction value. Existing work usually employ Pearson correlation coefficient (PCC) as the similarity measure. For example, the PCC similarity between user $u$ and user $v$ is defined as follows:
\begin{equation}\label{equ:sim1}
\small
sim(u, v) = \frac{\sum_{s \in J}{(R_{us} - \bar{R}_u)}{(R_{vs}
- \bar{R}_v)}}{\sqrt{\sum_{s \in J}(R_{us} -
\bar{R}_u)^2}{\sqrt{\sum_{s \in J}(R_{vs} -
\bar{R}_v)^2}}}~,
\vspace{-0.5ex}
\end{equation}
where $J = I_u \cap I_v$ is the set of Web services that are invoked by both user $u$ and user $v$. $R_{us}$ is the true QoS value of user $u$ invoking service $s$. $\bar R_u$ and $\bar R_v$ are the average QoS values observed by user $u$ and user $v$, respectively. From this definition, we have $sim(u, v) \in [-1, 1]$, where a larger PCC value indicates higher user similarity.

However, due to the obfuscation of QoS data, at server side we only have obfuscated QoS value $r'_{us}$, rather than its true value $R_{us}$. Therefore, we consider to employ $r'_{us}$ to approximately compute the similarity value $sim(u, v)$ as follows: 
\begin{eqnarray}
\small
&&\hspace{-8ex} sim(u, v) = \sum\nolimits_{s \in I_u \cap I_v}r'_{us}r'_{vs} / \sqrt{|I_u||I_v|} \label{equ:sim2} \\
\small
&&\hspace{-5ex}\approx~~ \sum\nolimits_{s \in I_u \cap I_v}r_{us}r_{vs} / \sqrt{|I_u||I_v|} \label{equ:sim3}\\
\small
&&\hspace{-5ex}=~~ \frac{\sum_{s \in I_u \cap I_v}{(R_{us} - \bar{R}_u)}{(R_{vs}
- \bar{R}_v)}}{\sigma_u\sigma_v\sqrt{|I_u||I_v|}} \label{equ:sim4}\\ 
\small
&&\hspace{-5ex}=~~ \frac{\sum_{s \in I_u \cap I_v}{(R_{us} - \bar{R}_u)}{(R_{vs}
- \bar{R}_v)}}{\sqrt{\sum_{s \in I_u}(R_{us} -
\bar{R}_u)^2}{\sqrt{\sum_{s \in I_v}(R_{vs} -
\bar{R}_v)^2}}}. \label{equ:sim5}
\vspace{-1ex}
\end{eqnarray}
By applying the scalar product property in Equ. (\ref{equ:scalarproduct}) to Equ. (\ref{equ:sim2}), substituting Equ. (\ref{equ:z-score}) to Equ.(\ref{equ:sim3}), and substituting Equ. (\ref{equ:sigma}) to Equ. (\ref{equ:sim4}), we derive Equ. (\ref{equ:sim5}), which is exactly the similarity measure used for collaborative filtering in the related work~\cite{PolatD03, HerlockerKBR99}. Note that this similarity measure differs slightly from Equ. (\ref{equ:sim1}) in the denominator part, but provides a good approximation to it (as the experiments shown in Section~\ref{sec:experiment}). Therefore, by using the obfuscated QoS data, we employ Equ. (\ref{equ:sim2}) as the approximation of the similarity between user $u$ and $v$.

After similarity computation between users, we can identify a set of top-k similar neighbours ($T_u$) for each user $u$. Then the unknown QoS value, for each entry where $r'_{us} = 0$, can be estimated as the weighted average of the QoS values observed by similar neighbours, i.e., 
\begin{equation}\label{equ:uipcc1}
\small
\hat r_{us}^U = \sum\nolimits_{v \in T_u}sim(u,v)r'_{vs} \Big/ \sum\nolimits_{v \in T_u}sim(u,v).
\vspace{-1ex}
\end{equation}

In a similar way, we can also leverage the information of similar services to make QoS prediction:
\begin{equation}\label{equ:uipcc3}
\small
\hat r_{us}^S = \sum\nolimits_{g \in T_s}sim(s,g)r'_{ug} \Big/ \sum\nolimits_{g \in T_s}sim(s,g),
\vspace{-1ex}
\end{equation}
where $T_s$ is the set of top-k similar services of service $s$. The similarity $sim(s,g)$ is further calculated by employing the cosine similarity between service $s$ and service $g$:
\begin{equation}\label{equ:uipcc2}
\small
sim(s, g) = \frac{\sum_{u \in I_s \cap I_g}{r'_{us}r'_{ug}}}{\sqrt{\sum_{u \in I_s \cap I_g}(r'_{us})^2}{\sqrt{\sum_{u \in I_s \cap I_g}(r'_{ug})^2}}}~,
\vspace{-0.5ex}
\end{equation}
where $I_s \cap I_g$ represents the set of users that invoke both service $s$ and service $g$. Note that the cosine similarity here equals to the original PCC similarity in UIPCC, because the QoS vectors have already been normalized during data obfuscation.

At last, as with UIPCC, a convex combination between user-based QoS prediction and service-based QoS prediction is employed to enhance the prediction accuracy.
\begin{equation}\label{equ:uipcc2}
\small
\hat r_{us} = \lambda\hat r_{us}^U + (1-\lambda)\hat r_{us}^S,
\vspace{-0.5ex}
\end{equation}
where $\lambda$ controls the combination weight between $\hat r_{us}^U$ and $\hat r_{us}^S$. Especially, when $\lambda=0$, $\hat r_{us} = \hat r_{us}^S$, and when $\lambda=1$, $\hat r_{us} = \hat r_{us}^U$.

However, this prediction result $\hat r_{us}$ is a normalized value that cannot reveal the prediction on the true QoS. When the user receives the prediction results from the server, a post-processing step, which is a re-normalization operation of the z-score normalization, can be taken to get the final prediction value $\hat R_{us}$:
\begin{equation}\label{equ:recover}
\small
\hat R_{us} = \bar R_u + \sigma_u * \hat r_{us}.
\vspace{-0.5ex}
\end{equation}
Note that the post-processing step can be only performed at user side because $\bar R_u$ and $\sigma_u$ are only known to the user.

\subsection{Privacy-Preserving PMF (P-PMF)}
PMF, or probabilistic matrix factorization~\cite{SalakhutdinovM07}, as a popular model-based collaborative filtering approach, has been suggested for QoS prediction by prior work~\cite{LoYDLW12, ZhengMLK13}. PMF works on an essential assumption of the low-rank structure of the QoS matrix. A matrix has a low rank when the entries of the matrix are largely correlated. In our case, as reported by the related work~\cite{ZhengMLK11, ZhengMLK13}, similar users usually have similar QoS values on the same Web service. The goal of PMF is to map $n$ users and $m$ services into a joint latent factor space with dimensionality $d$ such that each observed entry of the QoS matrix can be captured as the inner product of the corresponding latent factors.

Formally, we denote the latent user factors as $U \in \mathbb{R}^{d \times n}$ whose $u$-th column represents the latent factor of user $u$, and the latent service factors as $S \in \mathbb{R}^{d \times m}$ whose $s$-th column represents the latent factor of service $s$. Accordingly, we use $U_u^TS_s$ to approximate the observed QoS value $R_{us}$ between user $u$ and service $s$, i.e., $R_{us} \approx U_u^TS_s$, or more precisely,
\begin{equation}\label{equ:PMF1}
\small
R_{us} = U_u^TS_s + \delta_{us},
\vspace{-0.5ex}
\end{equation}
where $U_u^T$ is the transpose of $U_u$ and $\delta_{us}$ denotes the approximation error. The goal is to minimize all of the approximation errors. By taking $\delta_{us}$ as Gaussian noise~\cite{SalakhutdinovM07}, the loss function can be formulated as follows:
\begin{equation}\label{equ:PMF2}
\small
\mathcal{L} = \frac{1}{2}\sum\limits_{u = 1}^n \sum\limits_{s = 1}^m I_{us}{{(R_{us} - U^T_u{S_s})}^2} + \frac{\gamma}{2}(\sum_{u=1}^n\left\| U_u \right\|^2 + \sum_{s=1}^m\left\| S_s \right\|^2).
\vspace{-0.5ex}
\end{equation}
The first part measures the sum of squared approximation errors between $R_{us}$ and $U_u^TS_s$, where $I_{us}$ acts as an indicator that equals to 1 if $R_{us}$ is observed, and 0 otherwise. The second part are regularization terms used to avoid the overfitting problem, where ${\left\|  \cdot  \right\|}$ denotes the Euclidean norm, and $\gamma$ is a parameter to control the extent of regularization.

According to the basic PMF model as specified in Equ. (\ref{equ:PMF1}), the specific QoS of user $u$ invoking service $s$ can be effectively captured by the interaction between $U_u$ and service $S_s$. However, some other effects known as biases for determining the QoS values are independent of user-service interactions. For example, the users with high network bandwidth tend to experience fast network connections and the services equipped with abundant system resources likely provide short request-processing time. To capture these factors associated with either users or services, there is a suggestion for biased matrix factorization model in~\cite{YuLXY14}:
\begin{equation}\label{equ:P-PMF1}
\small
R_{us} = \mu + b_u + b_s + U_u^TS_s + \delta_{us},
\vspace{-0.5ex}
\end{equation}
where $\mu$ is a global bias, and $b_u$ and $b_s$ measure the user bias and service bias respectively. 

While preserving user privacy, the application of data obfuscation poses new challenges in modelling the obfuscated QoS data. To compromise the effect of data obfuscation, we set $\mu = 0$ and $b_u = \bar R_u$. Accordingly, we derive the following model:
\begin{equation}\label{equ:P-PMF2}
\small
r'_{us} = b'_s + {U'_u}^TS'_s + \delta'_{us} + \epsilon_{us}.
\vspace{-0.5ex}
\end{equation}
For ease of presentation, we further denote it as:
\begin{equation}\label{equ:P-PMF3}
\small
r'_{us} = b_s + U_u^TS_s + \delta_{us} + \epsilon_{us}.
\vspace{-0.5ex}
\end{equation}
This model naturally compromise the effect of z-score normalization at user side. By taking both $\delta_{us}$ and $\epsilon_{us}$ as Gaussian noise~\cite{SalakhutdinovM07}, the loss function can be expressed as:
\begin{eqnarray}\label{equ:P-PMF4}
\small
\mathcal{L}' &=& \frac{1}{2}\sum\limits_{u = 1}^n \sum\limits_{s = 1}^m I_{us}{{(r'_{us} - b_s - U^T_u{S_s})}^2} \nonumber \\
\small
&+& \frac{\gamma}{2}(\sum_{u=1}^n{b_s}^2 + \sum_{u=1}^n\left\| U_u \right\|^2 + \sum_{s=1}^m\left\| S_s \right\|^2).
\vspace{-0.5ex}
\end{eqnarray}
The minimization of this loss function can typically be solved by the gradient descent algorithm used in~\cite{ZhengMLK13} or the stochastic gradient descent algorithm used in~\cite{YuLXY14}. Due to space limits, we omit the algorithmic description here and refer interested readers to our supplementary report (see our project page). After obtaining the solutions with respect to $b_s$, $U_u$, and $S_s$, we can make the following QoS prediction:
\begin{equation}\label{equ:P-PMF5}
\small
\hat r_{us} = b_s + U^T_u{S_s}.
\vspace{-1ex}
\end{equation}

At last, as with P-UIPCC, a post-processing step in Equ.~(\ref{equ:recover}) is required to recover the prediction result $\hat r_{us}$ to the true prediction value $\hat R_{us}$. For both P-UIPCC and P-PMF, after obtaining the predicted QoS values of all the available Web services, we can recommend to users those services with top-ranked QoS values.

\begin{table}[!t]
\centering
\caption{\hspace{-2ex}Statistics of QoS Data}\label{tab:dataset} 
\vspace{0ex}
\begin{tabular}{c||c|c|c|c|c}
\hline
{QoS}& \#Users & \#Services & Range & Average & Std. \\
\hline 
 RT ($sec$)  & 339 & 5,825 & $0\sim20$ & $0.909$ & $1.973$ \\
\hline
 TP ($kbps$)  & 339 & 5,825 & $0\sim1000$ & $47.562$ & $110.797$ \\
\hline
\end{tabular}
\vspace{-2ex}
\end{table}

\section{Evaluation}\label{sec:experiment}
This section describes the experiments and the corresponding results of evaluating our privacy-preserving QoS prediction approaches. In particular, we intend to answer the following research questions.
\begin{itemize} \setlength{\itemindent}{0.5ex}
\setlength{\itemsep}{0.5ex}  
\item[RQ1:] \hspace{-0.5ex}What is the effect of data obfuscation?
\item[RQ2:] \hspace{-0.5ex}What is the accuracy of P-UIPCC and P-PMF?
\item[{RQ3}:] \hspace{-0.5ex}What is the tradeoff between accuracy and privacy?
\item[{RQ4}:] \hspace{-0.5ex}What is the effect of distribution of random noises on prediction accuracy?
\end{itemize}

\subsection{Experimental Setup}
In our experiments, we focus mainly on two representative QoS attributes: response time (RT) and throughput (TP).   
Response time measures the time duration between user sending out a
request and receiving a response, while throughput stands for the
data transmission rate of a user invoking a service.

The experiments are conducted based on a publicly-available QoS dataset of real-world Web services~\cite{ZhengZL14}. The dataset was collected in August 2009, providing a total of 1,974,675 response time and throughput records of service invocations between 339 users and 5,825 Web services. The 339 users are simulated by PlanetLab\footnote{PlanetLab (\url{https://www.planet-lab.org}) is an open platform for system and networking research, currently consisting of 1341 nodes at 654 global sites.} nodes distributed at 30 countries, while the 5,825 real-world Web services are crawled from the Internet and are deployed at 73 countries. Table~\ref{tab:dataset} provides a summary of the statistics of the data. 

In our experiments, we represent each type of QoS data by a 339-by-5825 QoS matrix with each entry denoting the observed response time/throughput of a specific invocation. In practice, the QoS matrix is very sparse because each user usually invokes only a handful of services. To simulate such data sparsity in our experiments, we randomly
remove entries from the full data matrix and only keep a small density of historical QoS values. Data density = 10\%, for example, indicates that each user invokes
10\% of the services, or each service is invoked by 10\% of
the users. We leverage the preserved data entries for QoS
prediction, and then use the removed QoS values as testing
data for accuracy evaluation.

To quantize the accuracy of QoS prediction, we employ a standard error metric, MAE (Mean Absolute Error), which has been widely used in the existing work (e.g.,~\cite{LoYDLW12, ZhengMLK13}).:
\begin{equation}
\small
MAE = {{{\sum\nolimits_{I_{us}=0} {\big| {\hat{R}_{us} - {R_{us}}} \big|} } \big{/}N}}~,
\vspace{-0.5ex}
\end{equation} 
where $R_{us}$ and $\hat{R}_{us}$ denote the observed QoS value and the corresponding predicted QoS value of the invocation between user $u$ and service $s$. $N$ is the total number of testing samples to be predicted, i.e., entries with $I_{us} = 0$. A smaller MAE value indicates better prediction accuracy.

\begin{figure}[!t]
    \centering
    \subfigure[$\alpha = 0$]{
        \includegraphics[width=0.15
        \textwidth]{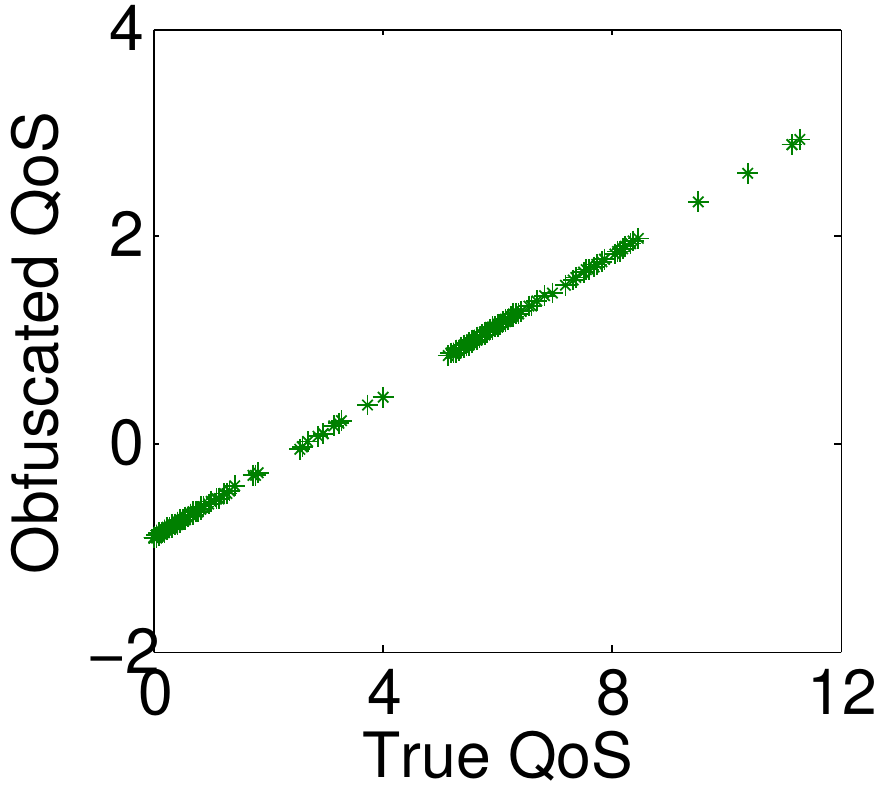}} 
        \hspace{-1ex}
    \subfigure[$\alpha = 0.5$]{
        \includegraphics[width=0.15
        \textwidth]{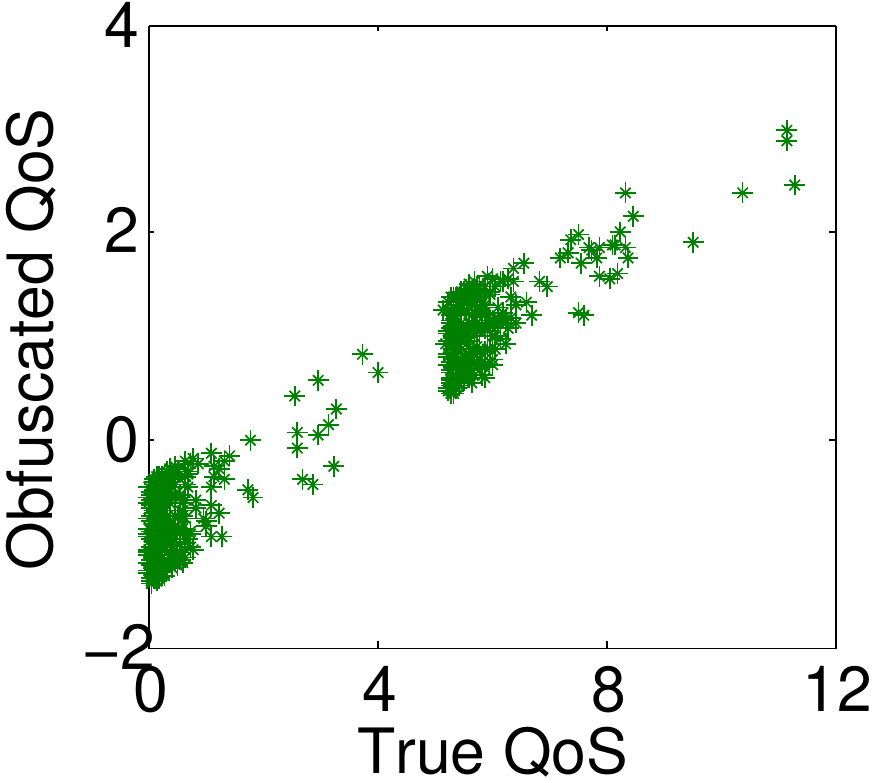}} 
        \hspace{-1ex}
    \subfigure[$\alpha = 1$]{
        \includegraphics[width=0.15
        \textwidth]{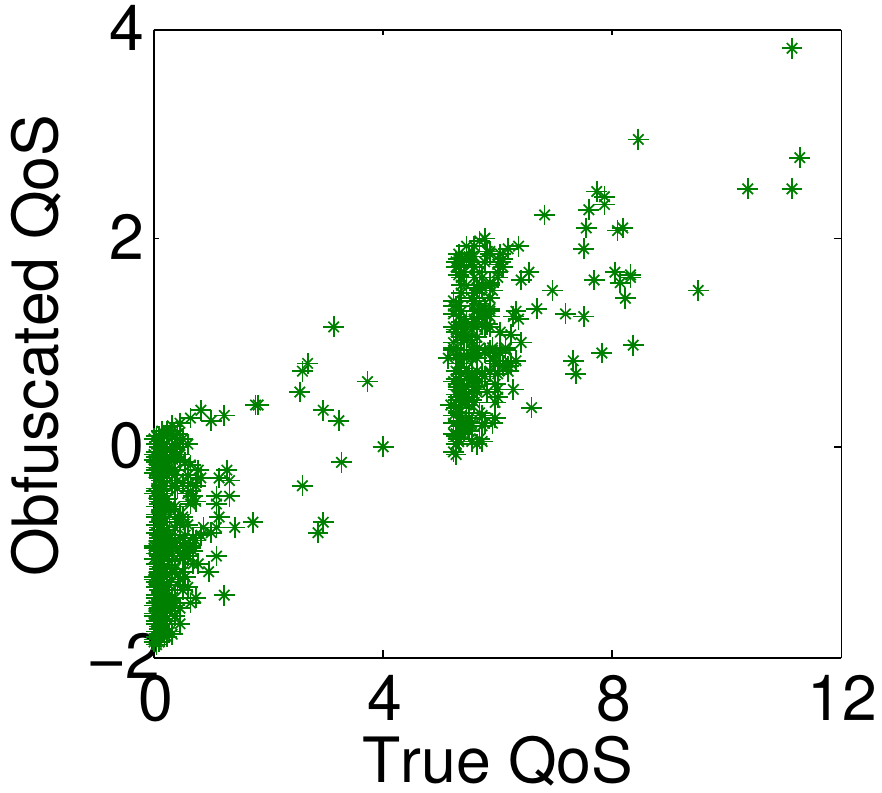}} 
    \vspace{-1ex}
  	\caption{Obfuscated QoS ($r'_{us}$) v.s. True QoS ($R_{us}$)}\label{fig:effectobfuscation}
  	\vspace{-2ex}
\end{figure}
\subsection{Effect of Data Obfuscation (RQ1)}
The aim of data obfuscation is to perturb the QoS data such that user privacy regarding the true QoS values can be preserved when performing collaborative analysis on the server. To understand the effect of data obfuscation made on QoS data (\textit{RQ1}), we compare the obfuscated QoS ($r'_{us}$) against the corresponding true QoS data ($R_{us}$). As an example, we randomly select a user from our dataset and provide three scatter plots by using the response time data of this user. The plots present the relationships between $r'_{us}$ and $R_{us}$ under different $\alpha$ settings. $\alpha$ is a parameter to determine the range of noises $\epsilon_{us}$ used to obfuscate the data. Especially, when $\alpha = 0$, the data obfuscation reduces to a z-score normalization process. Thus, Fig.~\ref{fig:effectobfuscation}(a) shows linear dependence between $r'_{us}$ and $R_{us}$. Z-score normalization is able to provide basic protection for user data where the mean and variance properties of QoS data are eliminated. The data after z-score normalization have a zero mean and unit variance. As $\alpha$ increases, the obfuscated data become more and more disordered, As shown in \ref{fig:effectobfuscation}(a) and (b), the linear correlation between $r'_{us}$ and $R_{us}$ is further eliminated. Consequently, a larger $\alpha$ indicates better protection for user data. Note that we have similar observations on the throughput data and thus omit the details here.

\subsection{Prediction Accuracy (RQ2)}
Data obfuscation is useful to perturb the QoS data for preserving user privacy, but it makes no sense without providing accurate prediction results. We evaluate the accuracy of our privacy-preserving QoS prediction approaches (P-UIPCC and P-PMF) based on the obfuscated QoS data, and compare them against the following baselines and counterpart approaches (\textit{RQ2}). We emphasize that these existing approaches require users' true QoS data and do not consider privacy issues.

\begin{itemize}
\setlength{\itemsep}{0.5ex}
\item \textbf{UMEAN}~\cite{ZhengMLK09}: This is a baseline approach that employs the average QoS value observed by a user (i.e., the row mean of $R$) to predict the unknown QoS of this user invoking other unused Web services. 
\item \textbf{IMEAN}~\cite{ZhengMLK09}: Likewise, this baseline approach employs the observed average QoS value of a Web service (i.e., the column mean of $R$) to predict the unknown QoS of other users invoking this Web service. 
\item \textbf{UIPCC}~\cite{ZhengMLK09, ZhengMLK11}: This is a hybrid approach that combines both user-based CF approach (UPCC) and item-based CF approach (IPCC) to make full use of the historical information from similar users and services for QoS prediction. UIPCC typically performs better than either UPCC or IPCC. 
\item \textbf{PMF}~\cite{ZhengMLK13}: This is a widely-used implementation of the matrix factorization model~\cite{SalakhutdinovM07}, which have been introduced to QoS prediction in~\cite{ZhengMLK13}.
\end{itemize}

\begin{table} [!t]
\centering \caption{\hspace{-2ex}Parameter Settings} \label{tab:settings}
\vspace{0ex}
\begin{tabular}{l|l|l|c|l|l|c}
\hline
Approach&\multicolumn{3}{c|}{RT} & \multicolumn{3}{c}{TP}\\
\hline
UIPCC   & $k:$ 10	& $\lambda:$ 0.1 	& -- & $k:$ 10 & $\lambda:$ 0.9	&  --   \\
P-UIPCC   &  $k:$ 10	&  $\lambda:$ 0.9	&  $\alpha:$ 0.5 & $k:$ 10 & $\lambda:$ 0.9	&  $\alpha:$ 0.5  \\
\hline
PMF   &  $d:$ 10	&  $\gamma:$ 40	& -- & $d:$ 10 & $\gamma:$ 800 	&  --  \\
P-PMF   & $d:$ 10 	&  $\gamma:$ 12	& $\alpha:$ 0.5 &$d:$ 10  & $\gamma:$ 12	&  $\alpha:$ 0.5  \\
\hline
\end{tabular}
\vspace{-1ex}
\end{table}

\begin{table} [!t]
\centering \caption{\hspace{-2ex}Prediction Accuracy (w.r.t. MAE)} \label{tab:comparison}
\vspace{0ex}
\begin{tabular}{c|l||c|c|c|c|c}
\hline
&&\multicolumn{5}{c}{Data Density}\\
\cline{3-7}
\raisebox{1.5ex}[0pt]{QoS} & \raisebox{1.5ex}[0pt]{Approach} &   10\%  &   15\%   & 20\% &  25\% & 30\%  \\
\hline
\hline
&   UMEAN   &  0.875 	&0.875 	&0.875 	&0.875 	&0.875     \\
&   IMEAN  &   0.688 	&0.683 	&0.681 	&0.680 	&0.679     \\
&   UIPCC  &  0.582 	&0.501 	&0.450 	&0.427 	&0.411  \\
&   PMF  &  0.487 	&0.452 	&0.431 	&0.418 	&0.409   \\
\cline{2-7}
&   P-UIPCC  &  0.569  & 0.537	  & 	0.512  & 0.495	 & 	0.482	 \\
\raisebox{6ex}[0pt]{RT} &   P-PMF  & 0.540   & 0.504	  & 0.478	 &0.458  & 	0.443	 \\
\hline
\hline
&   UMEAN   &   53.835 	&53.816 	&53.801 	&53.804 	&53.799     \\
&   IMEAN  &   26.860 	&26.716 	&26.641 	&26.593 	&26.571     \\
&   UIPCC  &  22.370 	&20.219 	&18.928 	&17.891 	&17.080   \\
&   PMF  &  15.994 	&14.670 	&13.924 	&13.405 	&13.117   \\
\cline{2-7}
&   P-UIPCC  &  23.572  & 	21.324 & 19.754 & 18.681	  & 	17.953	 \\
\raisebox{6ex}[0pt]{TP} &   P-PMF  & 20.702 & 18.451	  & 17.351	 & 	16.634  & 16.063	 \\
\hline
\hline
\end{tabular}
\vspace{-2ex}
\end{table}

For fair comparisons, we use the original parameters for the counterpart approaches, as specified in the related work, because we experiment on the same dataset. To make it consistent with these settings, most parameters of our approaches are set the same with them (e.g., $k=10$ for top-k neighbours in UIPCC and P-UIPCC). However, since both P-UIPCC and P-PMF work on obfuscated (normalized) data, we set different $\lambda$ and $\gamma$ values. The detailed parameters are specified in Table~\ref{tab:settings}. We use $\alpha=0.5$ in this experiment and study the effect of $\alpha$ in Section~\ref{sec:accuracyandprivacy}. Additionally, we vary the data density from 10\% to 30\% at a step increase of 5\%. Each approach is performed 20 times under each data density (with different random seeds), and the average MAE results are reported. 

Table~\ref{tab:comparison} provides the results of prediction accuracy with comparisons among different approaches. The results show that, while both of our approaches preserve decent privacy by data obfuscation ($\alpha$ = $0.5$), they still perform much better than the baselines including UMEAN and IMEAN, and achieve comparable accuracy with the counterpart approaches including UIPCC and PMF. In particular, P-UIPCC sometimes performs better than UIPCC (e.g., $0.569$ vs $0.582$), which can be attributed to the use of z-score normalization. Moreover, we observe that even working on obfuscated data, P-PMF mostly performs better than UIPCC. These encouraging results indicate the effectiveness of privacy-preserving approaches. In addition, we can see that the accuracy of these QoS prediction approaches improves with the increase in data density.

\subsection{Tradeoff between Accuracy and Privacy (RQ3)}\label{sec:accuracyandprivacy}
Whereas the goal of our work is to achieve both accuracy and privacy, there is indeed a tradeoff between them. At one extreme, users can provide true QoS data to obtain the most accurate QoS prediction results yet they lose privacy. At another extreme, users can submit totally false QoS data to preserve privacy but bad prediction results will be returned. To study such tradeoff between accuracy and privacy (\textit{RQ3}), we consider the effect of noise range $\alpha$ on prediction accuracy, because a larger $\alpha$ indicates better protection of privacy. Specifically, in this experiment, we set data density = 10\% and vary $\alpha$ from $0$ to $1$ at a step increase of $0.1$. Accordingly, we obtain the prediction accuracy under each $\alpha$ value. 

\begin{figure}[!t]
    \centering
    \subfigure[Response Time]{
        \includegraphics[width=0.22
        \textwidth]{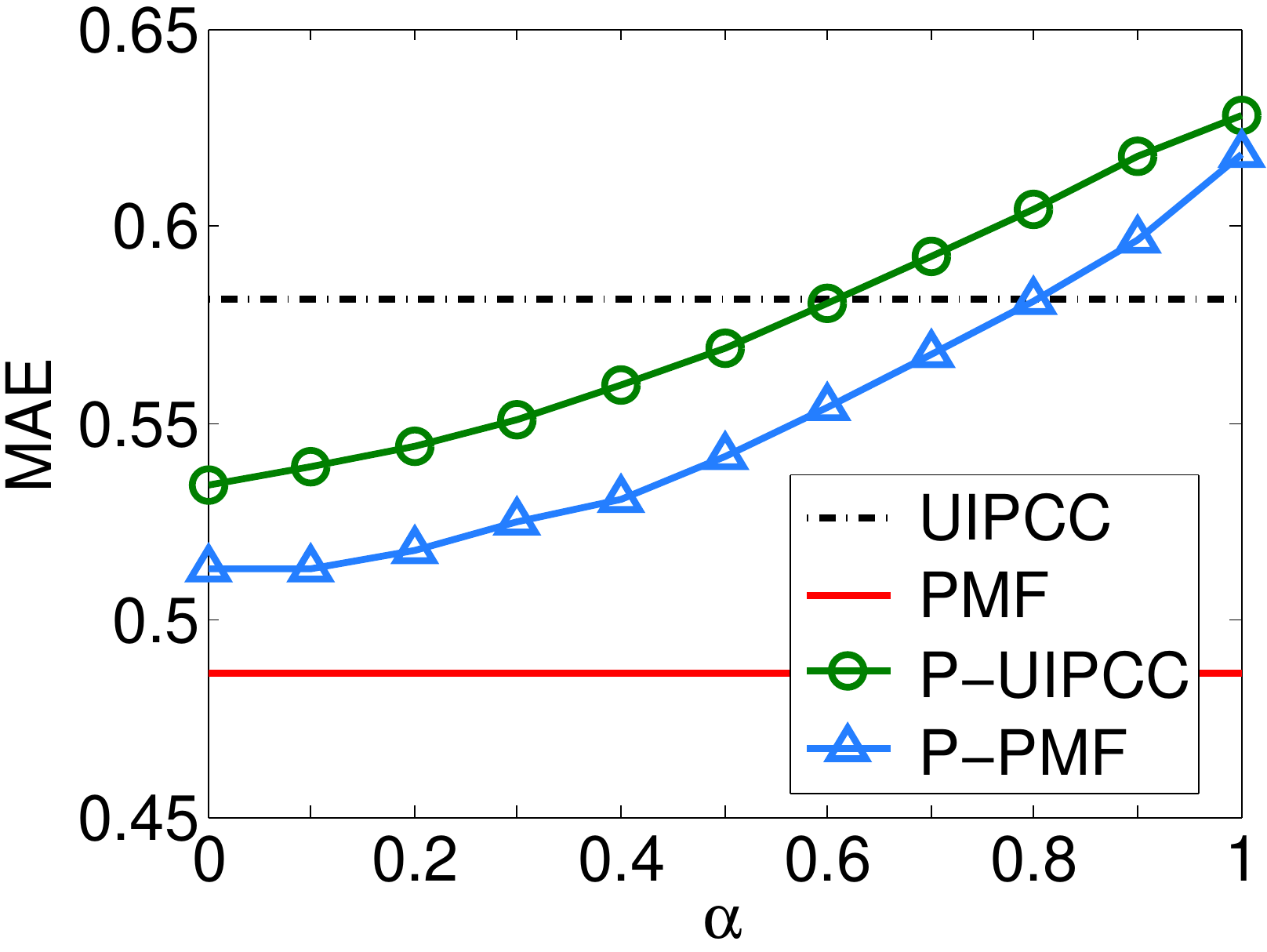}} 
        \hspace{1ex}
    \subfigure[Throughput]{
        \includegraphics[width=0.22
        \textwidth]{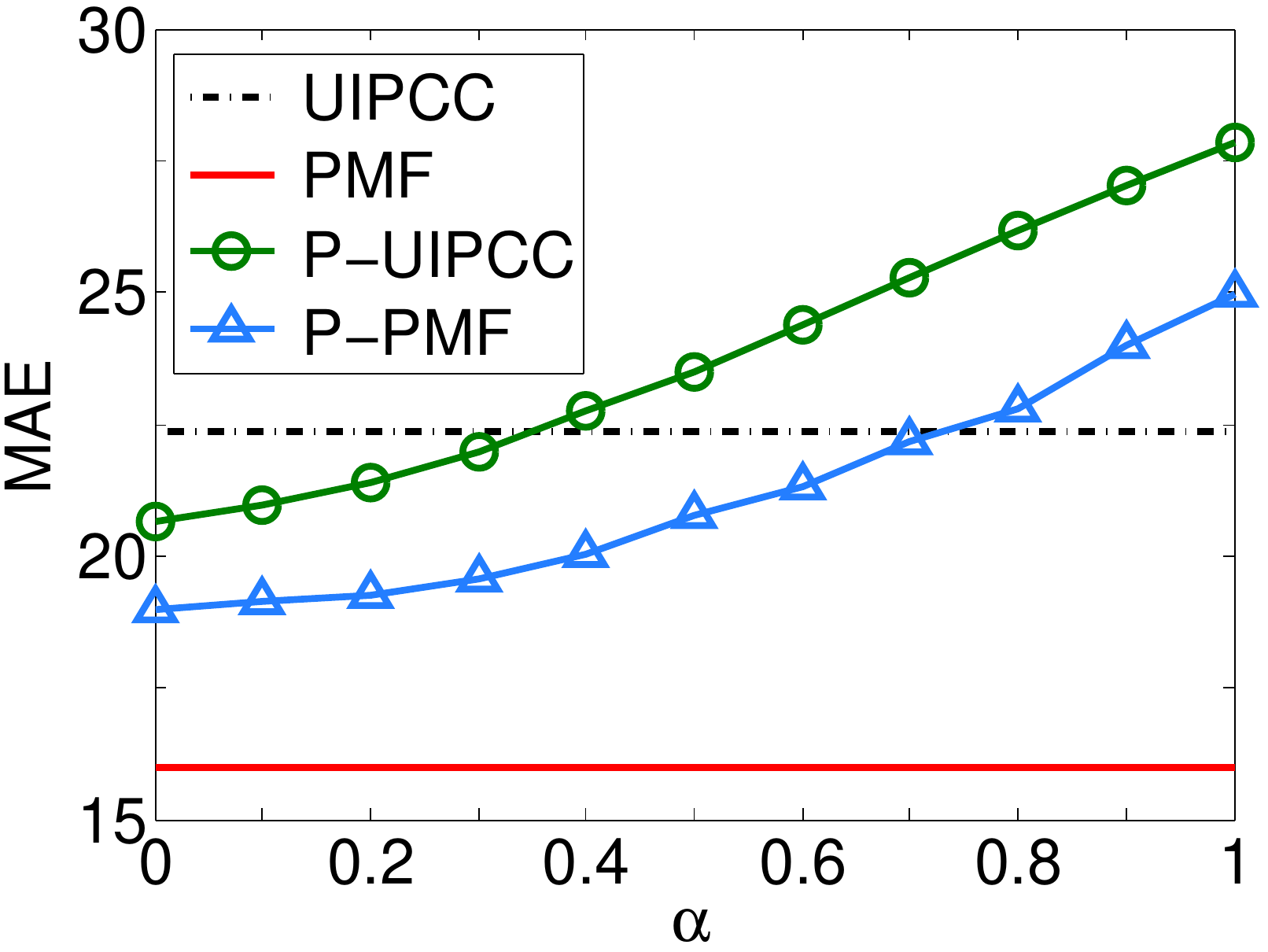}} 
    \vspace{-1ex}
  	\caption{Tradeoff between Accuracy and Privacy}\label{fig:tradeoffaccuracyandprivacy}
  	\vspace{0ex}
\end{figure}

Fig.~\ref{fig:tradeoffaccuracyandprivacy} presents the experimental results corresponding to response time and throughput, respectively. We can observe that both P-UIPCC and P-PMF degrade in accuracy (i.e., MAE increases) when $\alpha$ becomes larger, because the utility of data is less preserved. However, when $\alpha$ is small, e.g., less than 0.6 in Fig.~\ref{fig:tradeoffaccuracyandprivacy}(a), our privacy-preserving approaches are more accurate than UIPCC. Even $\alpha$ is as large as $1$, which is the variance of data after z-score normalization, the prediction accuracy is much better than the baselines (UMEAN and IMEAN). As a result, a balance needs to be made between the accuracy and privacy that a user wants to achieve. Additionally, we find that PMF and P-PMF consistently outperform UIPCC and P-UIPCC. This suggests the superior effectiveness of model-based approaches in capturing the latent structure of the QoS data, which conforms to the results reported in~\cite{ZhengMLK13}.

\begin{figure}[!t]
    \centering
    \subfigure[P-UIPCC]{
        \includegraphics[width=0.22
        \textwidth]{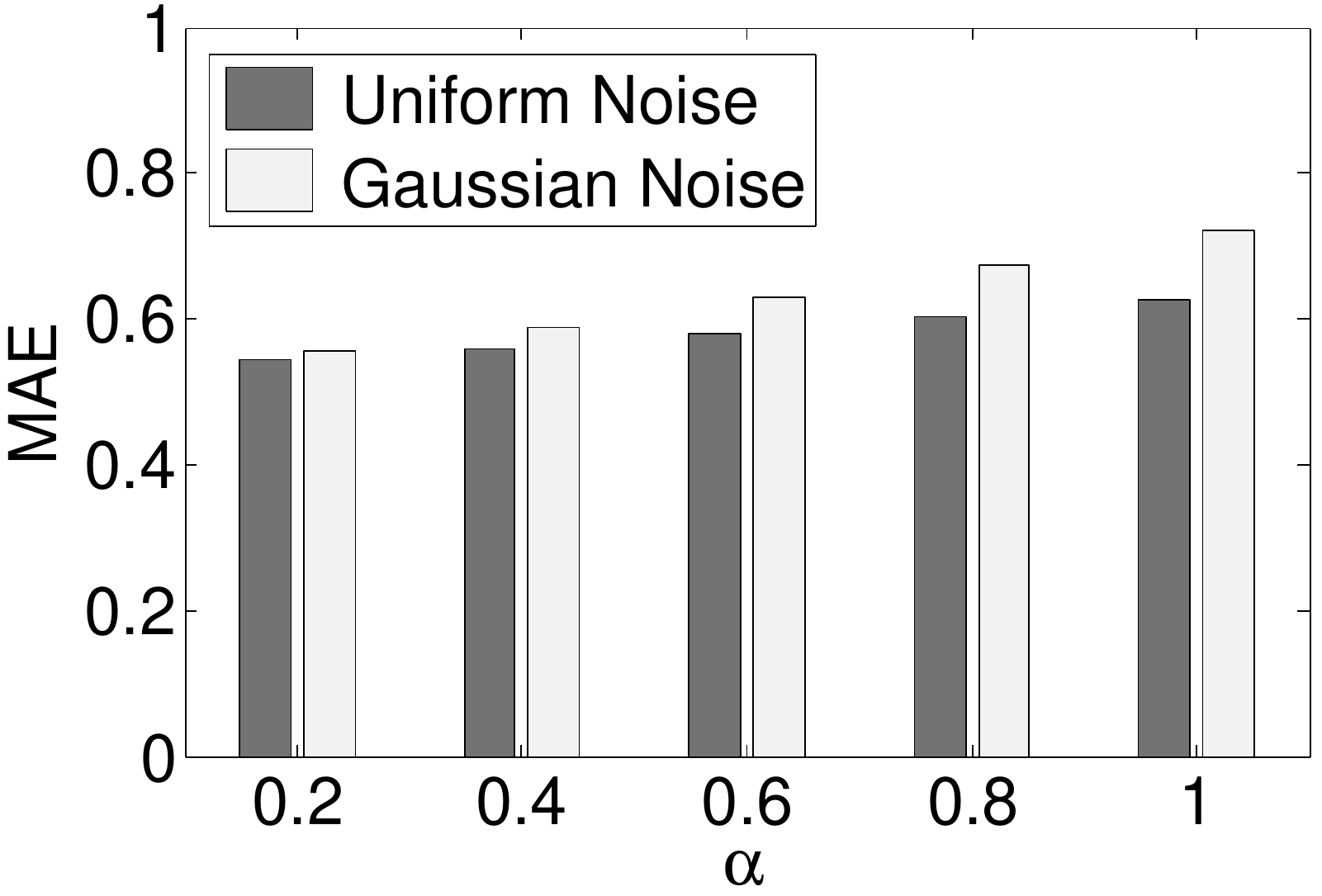}} 
        \hspace{1ex}
    \subfigure[P-PMF]{
        \includegraphics[width=0.22
        \textwidth]{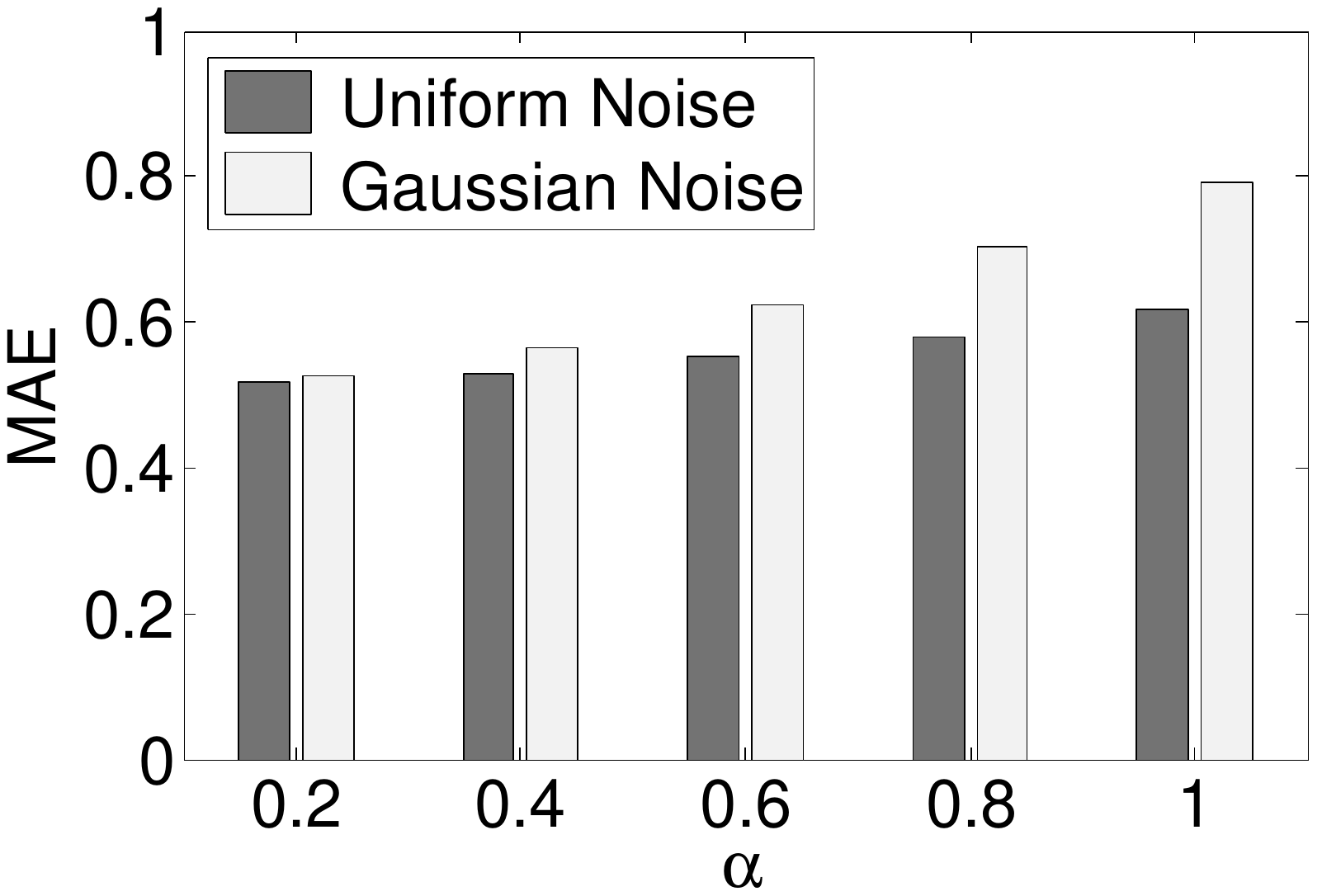}} 
    \vspace{-1ex}
  	\caption{Impact of Matrix Density}\label{fig:impactmatrixdensity}
  	\vspace{-2ex}
\end{figure}

\subsection{Effect of Distribution of Random Noises (RQ4)}\label{sec:impact_random}
In addition to the impact of noise range, a data randomization scheme is also subject to the choice of the distribution of random noises that are used for data obfuscation. In all of the above experiments, the random noises are generated from a uniform distribution located in [$-\alpha$, $\alpha$]. In contrast, in this experiment, we consider a Gaussian distribution $\mathcal{N}(0, \alpha)$ with a mean of zero and a standard deviation of $\alpha$. Compared to a uniform distribution, random noises generated from a Gaussian distribution are unevenly distributed. To investigate the effect of distribution of random noises (\textit{RQ4}), we vary the $\alpha$ value and compare the prediction accuracy of P-UIPCC and P-PMF with different settings on the distribution of random noises. 

Fig.~\ref{fig:impactmatrixdensity} presents the results of the accuracy comparison. We can observe that, for both P-UIPCC and P-PMF, the randomization scheme with uniform noises performs better than the scheme with Gaussian noises. In particular, the performance differs significantly between the two randomization schemes under a large $\alpha$ setting. The results imply that the distribution of random noises is a crucial factor for determining the performance of our privacy-preserving approaches.

\section{Conclusion}\label{sec:conclusion}
Privacy is a practical issue to be addressed for QoS-based Web service recommendation. This paper makes an initial effort to deal with the privacy-preserving Web service recommendation problem. We propose a generic privacy-preserving framework with the use of data obfuscation techniques, under which users can gain greater control on their data and rely less on the recommender system for privacy protection. We further develop two privacy-preserving QoS prediction approaches based on this framework, namely P-UIPCC and P-PMF, as representatives of neighbourhood-based CF approaches and model-based CF approaches respectively. To evaluate the effectiveness of P-UIPCC and P-PMF, we conduct experiments on a publicly-available QoS dataset of real-world Web services. The experimental results show that our privacy-preserving QoS prediction approaches can still descent prediction accuracy compared with the counterpart approaches. We hope that the encouraging results achieved in this initial work can inspire more research efforts on privacy-preserving Web service recommendation.

\section*{Acknowledgment}
The work described in this paper was supported by the
National Basic Research Program of China (973 Project No. 2011CB302603),
the National Natural Science Foundation of China (Project No. 61332010 and 61472338), and
the Research Grants Council of the Hong Kong Special Administrative
Region, China (No. CUHK 14205214 of the General Research Fund). 

\flushend
{\renewcommand\baselinestretch{0.88}\setlength{\parsep}{-0.15ex}\selectfont
\bibliographystyle{IEEEtran}
\bibliography{icws2015}
\par}


\end{document}